\documentstyle[12pt,epsfig]{article}
\begin{document}
\begin{center}
{\large \bf Low-temperature photoluminescence spectra\\ of highly
excited quantum wires}\\ {\bigskip Huynh Thanh Duc and D.B. Tran
Thoai}\\ {\bigskip \it Ho Chi Minh City Institute of Physics,\\
Vietnam Centre for Natural Science and Technology,\\ 1 Mac Dinh
Chi, Ho Chi Minh City, Viet Nam}\\
\end{center}
\begin{abstract}
Optical spectra of highly excited quantum wires at low
temperatures have been studied within the dynamically screening
approximation. We found a strong Fermi-edge singularity (FES) in
the photoluminescence spectra. The spectral shape and FES
intensity strongly depend on temperature in agreement with recent
experimental results.
\end{abstract}
\vspace{.5cm}

{\small \noindent {\it PACS:} 71.35.-y; 71.35.Ee; 73.23.Ps;
78.55.-m

Electron-electron interactions; D. Optical properties}

\newpage

\section{Introduction}
Many-body effects in dense electron-hole (e-h) plasma in
semiconductors and their heterostructures have found a growing
interest in the recent years. The interaction between the carriers
in dense e-h systems causes band-gap renormalization (BGR). The
band-gap decrease with increasing e-h pair densities affects the
excitation process and leads to optical nonlinearities [1-4].

Optical-nonlinear properties have been studied intensely in bulk
and low- dimensional semiconductors [1-12]. At high densities the
low-temperature optical spectra of semiconductors show a peak near
the Fermi energy. This peak has been known as the excitonic
enhancement \cite{Haug&Koch,Zimmermann1}. In semiconductor quantum
wires (QWRs), the screening effect is much weaker so that the
excitonic enhancement becomes very strong and one obtains the
Fermi-edge singularity (FES) \cite{klingshirn}. In their
experiments at low temperatures, Calleja et al. \cite{calleja1}
have found that the photoluminescence (PL) spectra of
modulation-doped GaAs quantum wires are dominated by a Fermi-edge
singularity, which shows a strong temperature dependence and a
pronounced spectra shape.

In this work, we calculate the low-temperature PL spectra of GaAs
QWRs by solving the Bethe-Salpeter equation (BSE) of the e-h pair
Green's function within the dynamically screened approximation.
Our calculation includes the following many-body effects:
phase-space filling, screening of the Coulomb interaction,
band-gap renormalization. Instead of using a phenomenological
damping constant we shall take into account the broadening and
lifetime effects by determining the imaginary part of the electron
and hole self-energy. We shall compare our theoretical results
with experimental findings by Calleja et al. in GaAs quantum wires
\cite{calleja1} and with theoretical results obtained from the
statically screened approximation and the Hartree-Fock
approximation as well.

\section{Theory}
We start from an ideal 2D quantum well in which the electrons and
holes are confined laterally by a harmonic oscillator potential
with a total intersubband spacing $\Omega$. The 1D Coulomb
interaction between the charge carriers is given by averaging the
bare 2D Coulomb interaction with the lateral envelope wave
functions:
\begin{equation}\label{bare}
V(q)=\frac{2e^2}{\epsilon_0}
\exp\left(\frac{q^2}{4m\Omega}\right)K_0\left(\frac{q^2}{4m\Omega}\right),
\end{equation}
where $m$ is the reduced electron-hole mass and $K_0(x)$ is the
zeroth-order modified Bessel function.

The photoluminescence spectrum $R(\omega)$ is calculated from the
imaginary part of the optical dielectric function
$\epsilon(\omega)$:
\begin{eqnarray}
R(\omega)\sim\frac{Im\epsilon(\omega)}{e^{\beta(\omega-\mu)}-1},
\end{eqnarray}
where $\beta=1/k_BT$ is the inverse thermal energy,
$\mu=\mu^e+\mu^h$ the combined chemical potential of the e-h pair.

In the linear response theory \cite{HSR} the optical dielectric
function of e-h system is obtained from:
\begin{eqnarray}
\epsilon(\omega)=\epsilon_\infty - 4\pi e^2
\sum_{k,k'}r_{vc}(k)r^*_{vc}(k')G^{}_{eh}(k,k',\omega).
\end{eqnarray}
where $G^{}_{eh}(k,k',\omega)$ is the e-h pair Green's function
[2,5,9-10].

The effective Bethe-Salpeter equation for the e-h pair Green's
function in the dynamically screened approximation is given by
[2,5,9-10]:
\begin{eqnarray}\label{bse}
G(k,k',\omega)=\frac{1-f^e(e^e_k)-f^h(e^h_{-k})}{\omega-e^e_k-e^h_{-k}}
\left(\delta_{k,k'}-\sum_{k''}V_{eff}(k,k'',\omega)G(k'',k',\omega)\right),
\end{eqnarray}
where $f^j(\varepsilon)=(e^{\beta(\varepsilon-\mu^j)}+1)^{-1}$ is
the Fermi distribution function,
\begin{eqnarray}
e^j_k=\frac{E_g^0}{2}+\frac{k^2}{2m^j}+\Sigma^j(k,e^j_k)
\end{eqnarray}
are the renormalized band energies for electrons $(j=e)$ and holes
$(j=h)$,
\begin{eqnarray}
V_{eff}(k,k',\omega) &=& \left(\frac{1}{\beta}\right)^2
\sum\limits_{z,z'}\frac{G^e(k,\Omega-z)+G^h(-k,z)}{1-f^e(e^e_k)-f^h(e^h_{-k})}
V_S(k-k',z-z')\nonumber\\ &&
\times\left.\frac{G^e(k',\Omega-z')+G^h(-k',z')}{1-f^e(e^e_{k'})-f^h(e^h_{-k'})}\right|_{\Omega=\omega-\mu^e-\mu^h}
\end{eqnarray}
is the effective e-h interaction, and $G^j(k,z)$ is the
single-particle Green's function.

In the plasmon-pole approximation (PPA) [1-3,5,9-11], the
dynamically screened Coulomb potential is given by:
\begin{eqnarray}\label{sc}
V_S(k-k',z-z') =
V(k-k')\left(1+\frac{\omega_{pl}^2(k-k')}{(z-z')^2-\omega_{k-k'}^2}\right),
\end{eqnarray}
where
\begin{eqnarray}
\omega_{pl}^2(q)&=&\sum\limits_{j=e,h}\frac{V(q)n^jq^2}{m^j},\nonumber\\
\omega_{q}^2&=&\omega^2_{pl}+\sum\limits_j\frac{n^jq^2}{\kappa
m^j}+\frac{q^4}{4m^2},\nonumber\\ \kappa
&=&2\beta\sum\limits_{j,k}f^j(e^j_k)(1-f^j(e^j_k)).
\end{eqnarray}
The effective e-h interaction in the PPA reads:
\begin{eqnarray}\label{veff}
V_{eff}(k,k',\omega)=V(k-k')\left(1+\zeta(k,k',\omega)\right),
\end{eqnarray}
where
\begin{eqnarray}
\zeta(k,k',\omega)&=&\frac{1}{1-f^e(e^e_k)-f^h(e^h_{-k})}\
\frac{1}{1-f^e(e^e_{k'})-f^h(e^h_{-k'})}\
\frac{\omega^2_{pl}(k-k')}{2\omega_{k-k'}}\nonumber\\
&\times&\left[\
\frac{(g(\omega_{k-k'})+f^e(e^e_{k'}))(f^e(e^e_{k'}-\omega_{k-k'})-f^e(e^e_{k'}))}{e^e_k-e^e_{k'}+\omega_{k-k'}}\right.\nonumber\\
&&+\
\frac{(1+g(\omega_{k-k'})-f^e(e^e_{k'}))(f^e(e^e_{k'}+\omega_{k-k'})-f^e(e^e_k))}{e^e_k-e^e_{k'}-\omega_{k-k'}}\nonumber\\
&&+\
\frac{(g(\omega_{k-k'})+f^h(e^h_{-k'}))(f^h(e^h_{-k'}-\omega_{k-k'})-f^h(e^h_{-k}))}{e^h_{-k}-e^h_{-k'}+\omega_{k-k'}}\nonumber\\
&&+\
\frac{(1+g(\omega_{k-k'})-f^h(e^h_{-k'}))(f^h(e^h_{-k'}+\omega_{k-k'})-f^h(e^h_{-k}))}{e^h_{-k}-e^h_{-k'}-\omega_{k-k'}}\nonumber\\
&&+\
\frac{(g(\omega_{k-k'})+f^e(e^e_{k'}))(1-f^e(e^e_{k'}-\omega_{k-k'})-f^h(e^h_{-k}))}{\omega-e^e_{k'}-e^h_{-k}+\omega_{k-k'}}\nonumber\\
&&+\
\frac{(1+g(\omega_{k-k'})-f^e(e^e_{k'}))(1-f^e(e^e_{k'}+\omega_{k-k'})-f^h(e^h_{-k}))}{\omega-e^e_{k'}-e^h_{-k}-\omega_{k-k'}}\nonumber\\
&&+\
\frac{(g(\omega_{k-k'})+f^h(e^h_{-k'}))(1-f^e(e^e_k)-f^h(e^h_{-k'}-\omega_{k-k'}))}{\omega-e^e_k-e^h_{-k'}+\omega_{k-k'}}\nonumber\\
&&+
\left.\frac{(1+g(\omega_{k-k'})-f^h(e^h_{-k'}))(1-f^e(e^e_k)-f^h(e^h_{-k'}+\omega_{k-k'}))}{\omega-e^e_k-e^h_{-k'}-\omega_{k-k'}}\right],
\end{eqnarray}
and $g(\omega)=(e^{\beta\omega}-1)^{-1}$ is the Bose distribution
function.

We calculate the self-energy $\Sigma^j(k,\omega)$ within the
leading-order approximation:
\begin{eqnarray}\label{se}
\Sigma^j(k,\omega) &=&\Sigma^j_{HF}(k)+\Sigma^j_{corr}(k,\omega),
\end{eqnarray}
where
\begin{eqnarray}
\Sigma^j_{HF}(k)&=& -\sum_{k'}V(k-k')f^j(e^j_{k'})
\end{eqnarray}
is the unscreened Hartree-Fock self-energy and
\begin{eqnarray}
\Sigma^j_{corr}(k,\omega)&=&\sum_{k'}V(k-k')\frac{\omega^2_{pl}(k-k')}{2\omega_{k-k'}}\nonumber\\
&\times&\left[\frac{1+g(\omega_{k-k'})-f^j(e^j_{k'})}{\omega-e^j_{k'}-\omega_{k-k'}+i\gamma}
+\frac{g(\omega_{k-k'})+f^j(e^j_{k'})}{\omega-e^j_{k'}+\omega_{k-k'}+i\gamma}\right].
\end{eqnarray}
is the correlation self-energy. $\gamma$ is a small
phenomenological damping term describing impurity, defect
scattering, inhomogeneities in the system, and other possible
broadening process. $\gamma$ should be small compared with the
excitonic binding energy in quantum wires \cite{Sarma}.

If one replaces the effective e-h interaction
$V_{eff}(k,k',\omega)$ in (\ref{veff}) by the statically screened
interaction $V_S(k-k')$ or the bare Coulomb potential $V(k-k')$
and the self-energy in (\ref{se}) by
$\Sigma^j(k)=\sum\limits_{k'}\left[-V_S(k-k')f^j(e^j_{k'})+\frac{1}{2}(V_S(k')-V(k'))\right]$
or $\Sigma^j(k)=\Sigma^j_{HF}(k)$, one retrieves the BSE in the
statically screened approximation treated by other authors
[1-3,11] or in the Hartree-Fock approximation treated by Gr\'{e}us
et al \cite{Greus}. Note that the self-energy in the statically
screened approximation or in the Hartree-Fock approxiamation is
real and one has to introduce a phenomenological damping constant
$\Gamma$, which is corresponding to the sum of the imaginary part
of the self-energy of electron and hole, to describe the
broadening and lifetime effects [1-3,6,11].

\section{Numerical results and discussions}
We calculate PL spectra by solving the BSE for GaAs QWR using the
following parameters: $m_e=0.067\ m_0$, $m_h=0.46\ m_0$,
$\epsilon_0=13.1$, $\Omega=5.2\ \mathrm{meV}$, $E_g^0=1.53\
\mathrm{eV}$, $\gamma=2\ {\mathrm{meV}}<< E_b$ ($E_b\sim 20\
{\mathrm{meV}}$ is excitonic binding energy in GaAs QWR).

In Fig. 1 we compare our PL spectra in the dynamically screening
approximation (Fig. 1a) with experimental results by Calleja et
al. (Fig. 1b) taken for e-h density $n=5\times 10^5\
{\mathrm{cm}^{-1}}$ and at different temperatures \cite{calleja1}.
At very low temperature ($T=1.7\ \mathrm{K}$) we obtain a strong
FES near the chemical potential $\mu$. With increasing
temperatures the peak shifts to the red and the FES intensity
decreases quickly due to increasing thermal broadening of the
Fermi function and increasing thermal collisions. Our theoretical
results agree rather well with the PL measured results (Fig. 1b)
in modulated-doped GaAs QWRs.

In Figs. 2 we plot PL spectra in the unscreened Hartree-Fock
approximation (Fig. 2a) and in the statically screened
approximation (Fig. 2b) at $n=5\times 10^5\ {\mathrm{cm}^{-1}}$
using a damping constant $\Gamma=4\ \mathrm{meV}$. In this
simplest Hartree-Fock approximation the PL intensity increases
suddenly when the chemical potential coincides with the excitonic
resonance energy (at $T\simeq 10\ \mathrm{K}$). Besides that, the
spectra show a blue shift with increasing temperatures. Contrary
to the Hartree-Fock approximation the statically screened
approximation shows a red shift with increasing temperatures (Fig.
2b) but the spectral shape is still much less pronounced compared
to that of measurements.

In summary, we have studied the temperature dependence of PL
spectra of photoexcited QWRs in three different approximations:
Hartree-Fock, statically screened-, and dynamically screened
approximation. While the dynamically screened approximation agrees
rather well with recent experimental findings \cite{calleja1} the
statically screened approximation poorly describes the
measurements. The unscreened Hartree-Fock approximation, moreover,
fails totally to describe the experiments.
\\

\noindent{\large\bf{Acknowledgements}}

\vspace{.5cm}\noindent We gratefully acknowledge the financial
support of the National Program for Basic Research.



\newpage
\epsfig{figure=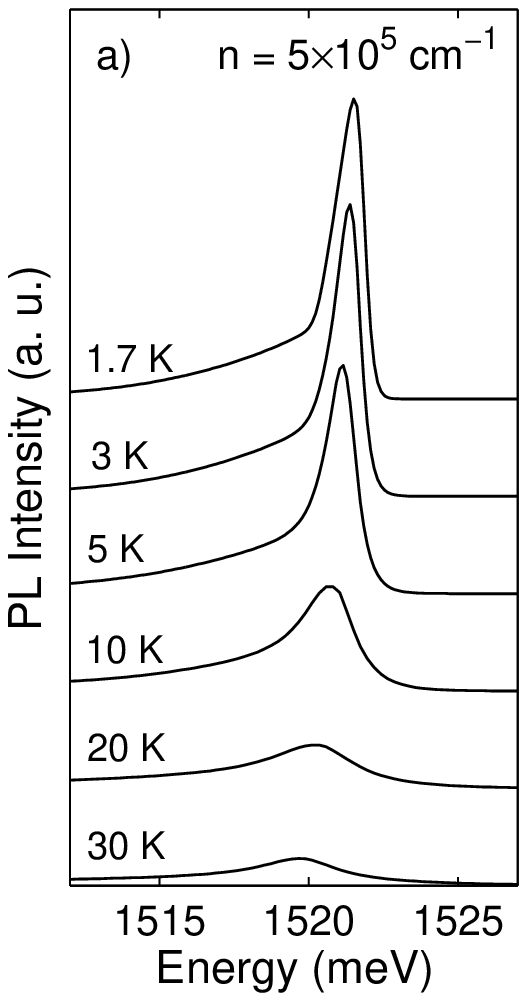, width=6cm}\ \ \ \ \
\epsfig{figure=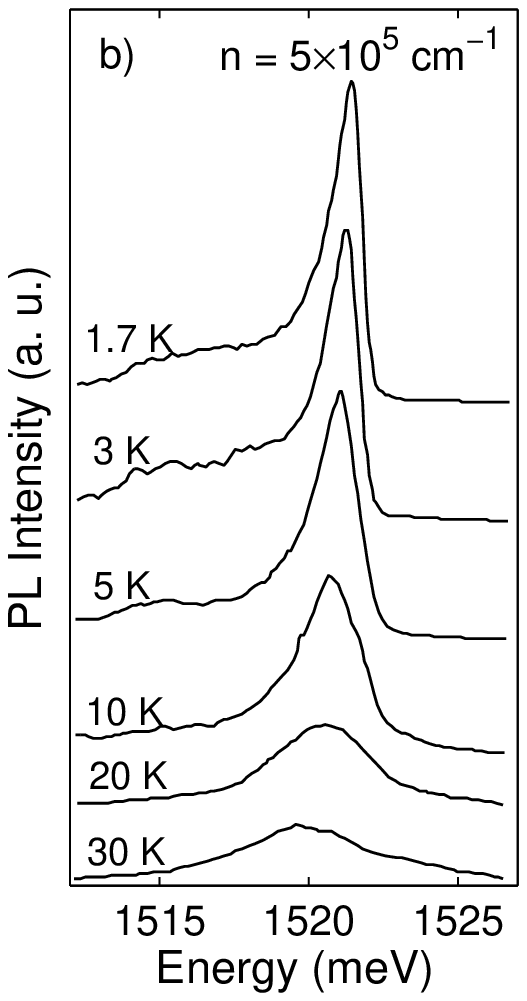, width=6cm}


\bigskip\noindent Figure 1a: PL spectra of GaAs QWR for various temperatures in the dynamically screening approximation.

\bigskip\noindent Figure 1b: Experimental PL spectra of GaAs QWRs \cite{calleja1} for various temperatures at evaluated intersubband spacing of $5.2\ \mathrm{meV}$, and plasma density of $5\times 10^5 \mathrm{cm^{-1}}$.

\newpage
\epsfig{figure=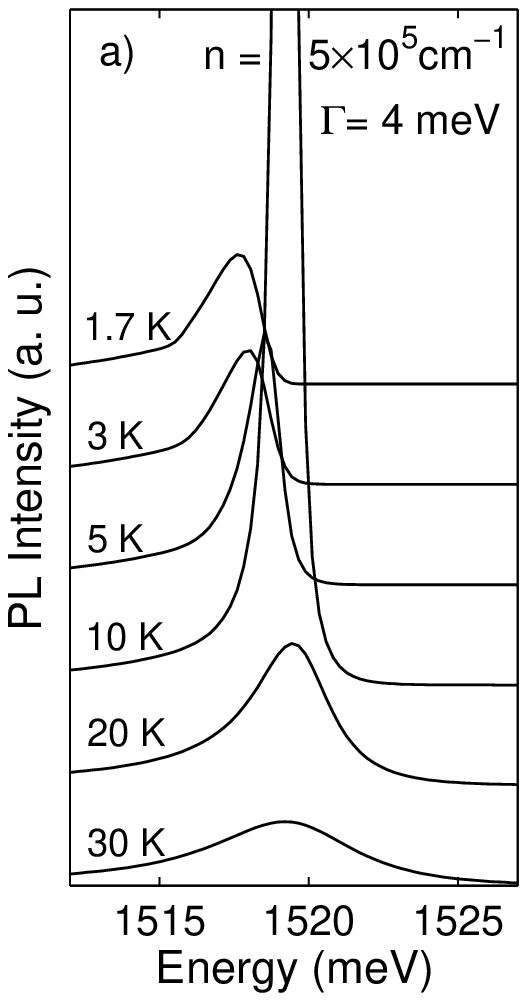, width=6cm}\ \ \ \ \
\epsfig{figure=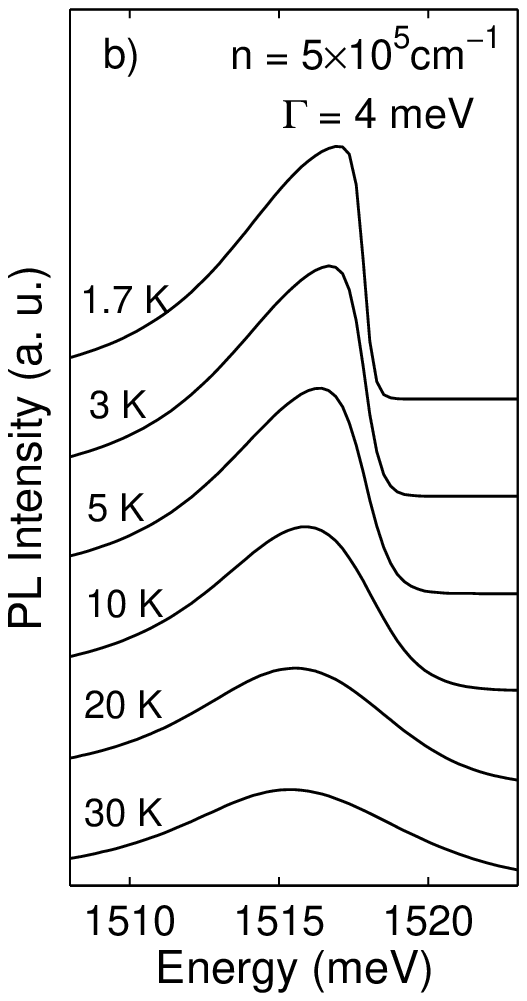, width=6cm}


\bigskip\noindent Figure 2a: PL spectra of the GaAs QWR for various temperatures in the Hartree-Fock approximation.

\bigskip\noindent Figure 2b: PL spectra of the GaAs QWR for various temperatures in the statically screened approximation.




\begin{thebibliography}{20}
\bibitem{Haug&Koch}H. Haug, S.W. Koch, {\it Quantum Theory of the
Optical and Electronic Properties of Semiconductors}, World
Scientific, Singapore (1993).
\bibitem{HSR}H. Haug, S. Schmitt-Rink, Prog. Quant. Electr. {\bf 9} (1984) 3.
\bibitem{Zimmermann1}R. Zimmermann, {\it Many Particle Theory of Highly Excited Semiconductors}, Teubner, Liepzig (1988).
\bibitem{klingshirn}C.F. Klingshirn, {\it Semiconductor Optics},
Springer-Verlag, Berlin 1995.
\bibitem{Haug}H. Haug, D.B. Tran Thoai, phys. stat. sol. (b) {\bf
85} (1978) 561.
\bibitem{Greus}Ch. Gr\'{e}us, A. Forchel, R. Spiegel, F. Faller, S. Benner, H. Haug, Europhys. Lett. {\bf 34} (1996) 213.
\bibitem{Rossi}F. Rossi, E. Molinari, Phys. Rev. Lett. {\bf 76}
(1996) 3642.
\bibitem{Ambigapathy}R. Ambigapathy, I. Bar-Joseph, D.Y. Oberli,
S. Haacke, M.J. Brasil, F. Reinhardt, E. Kapon, B. Deveaud, Phys.
Rev. Lett. {\bf 78} (1997) 3579.
\bibitem{Sarma1}S. Das Sarma, D.W. Wang, Phys. Rev. Lett. {\bf 84} (2000) 2010.
\bibitem{Sarma}D.W. Wang, S. Das Sarma, Phys. Rev. B {\bf 64} (2001) 195313.
\bibitem{duc}H.T. Duc, D.B. Tran Thoai, Solid State Commun. {\bf
118}
(2001) 211.
\bibitem{calleja1}J.M. Calleja, J.S. Weiner, A.R. Go\~{n}i, A. Pinczuk, {\it Optics of
Semiconductor Nanostructures}, edited by F. Henneberger, S.
Schmitt-Rink, E.O. Gobel, Akademie Verlag 1995, p. 335.
\end{thebibliography}
\end{document}